\documentclass[12pt]{article}
\usepackage{epsfig}
\usepackage{amsfonts}
\usepackage{amsmath}
\usepackage{amssymb}
\usepackage{amscd}
\usepackage{afterpage}
\usepackage{subeqnarray}
\usepackage{float}
\allowdisplaybreaks[4]

\begin{document}

\title{On the analytic solution of 
the pairing problem:\\ one pair in many levels}
\author{M. Barbaro$^1$, R. Cenni$^2$,
A. Molinari$^1$
and M. R. Quaglia$^2$\\
${}^1$ Dipartimento di Fisica Teorica --- Universit\`a di Torino
\\
Istituto Nazionale di Fisica Nucleare --- Sez. di Torino\\
 Torino ---Italy \\
${}^2$ Dipartimento di Fisica --- Universit\`a di Genova\\
Istituto Nazionale di Fisica Nucleare --- Sez. di Genova\\
Genova --- Italy\\}
\date{}
\maketitle

\begin{abstract}
We search for approximate, but analytic solutions of the pairing problem for 
one pair of nucleons in many levels of a potential well. 
For the collective energy a general formula,
independent of the details of the single particle spectrum, is given in both
the strong and weak coupling regimes.
Next the displacements of the solutions trapped in between the single
particle levels with respect to the unperturbed energies are explored: 
their dependence upon a suitably defined 
quantum number is found to undergo a transition between two different regimes.
\end{abstract}

PACS: 24.10.Cn;~~21.60.-n

Keywords: Pairing interaction

\section{Introduction}
\label{sec:intro}

In this paper we deal with the problem of the pairing Hamiltonian for one
pair of nucleons living in a set of levels of a potential well.

There are several motivations for this study. First, to find approximate, 
but analytic, solutions for the eigenvalues and the eigenfunctions, which we 
believe to be not only interesting {\it per se}, but also useful for paving the
way to the general problem of $n$ interacting pairs 
(see ref.~\cite{RiSh-64}). Second, to connect the 
energy of the collective mode, both in the strong and in the weak 
coupling regime, to global features of the single particle levels spectrum
like the variance, the skewness, the kurtosis, etc. 
Third, to unravel the remarkable pattern displayed by the solutions trapped in 
between the single particle levels, 
already hinted at in ref.~\cite{BaFoMoQu-01}.
Indeed by connecting the trapped solutions to a quantum
number $\lambda$, it is found that their behaviour versus $\lambda$ is not 
only smooth, but displays a transition between two different regimes.
Interestingly, this transition may be on one side related to a sort of
sum rule obeyed by the trapped eigenvalues (stemming from the Vi\`ete 
conditions for the solutions of an algebraic equation) and on the other to
the basic nature of the pairing interaction.

The paper is organised as follows:
in section~\ref{sec:form} 
the general formalism is presented and the basic equations
are deduced in the framework of the Grassmann variables;
in section~\ref{sec:coll} the collective solution is addressed in both the 
strong and weak coupling regimes; in section~\ref{sec:trapped} the trapped 
solutions and their regularities are studied within the harmonic oscillator
and related models for the 
single particle potential well.

\section{General formalism}
\label{sec:form}

Consider a shell of an average potential well, whose shape needs not to be 
specified, with $L$ non-degenerate single particle levels of energy $e_\nu$ 
and  angular momentum $j_\nu$, the associated multiplicity being
$2 \Omega_\nu = 2 j_\nu + 1$ ($1\leq \nu\leq L$).

Let identical fermions (e.g., neutrons) living in this set of levels
interact through the pairing force.
The Hamiltonian of the system 
$\hat H=\hat H_0+\hat H_P$
splits then into a single particle 
\begin{equation}
  \label{eq:x2}
  \hat H_0= \sum_{\nu=1}^L e_\nu \sum_{m_\nu=-j_\nu}^{j_\nu}
\hat a^\dagger_{j_\nu m_\nu} \hat a_{j_\nu m_\nu}
\end{equation}
and into a pairing interaction
\begin{equation}
  \label{eq:x3}
  \hat H_P= - G \sum_{\mu,\nu=1}^L \hat A^\dagger_\mu \hat A_\nu
\end{equation}
term.
In (\ref{eq:x3})
\begin{equation}
  \label{eq:x4}
  \hat A_\mu = \sum_{m_\mu=1/2}^{j_\mu} (-1)^{j_\mu-m_\mu}
\hat a_{j_\mu,- m_\mu} \hat a_{j_\mu m_\mu}\ ,
\end{equation}
$\hat a_{jm}$, $\hat a_{jm}^\dagger$ being the nucleon's
destruction and creation operators.
The operators $\hat A_\mu$ and $\hat A^\dagger_\mu$ destroy
and create pairs having total angular momentum $J=0$ 
in the level $j_\mu$.

As well-known, the problem of the Hamiltonian (\ref{eq:x2},\ref{eq:x3})
has been addressed \cite{RiSh-64} by first diagonalising the associated 
bosonic Hamiltonian and then by accounting for the Pauli principle.
Here we search for compact, possibly accurate,
expressions for the eigenvalues and eigenvectors of $\hat H$ and to 
unravel hidden correlations among the solutions.

We start by recasting the eigenvalue equation
in the Bargmann-Fock representation of the hamiltonian formalism,
where the odd (anti-commuting) Grassmann variables $\lambda_{jm}$,
$\lambda^*_{jm}$ replace the fermionic operators
$\hat a_{jm}$, $\hat a_{jm}^\dagger$.
For this purpose we introduce the even  variables
\begin{equation}
  \label{eq:x5}
  \varphi_{j m} \equiv (-1)^{j-m} \lambda_{j -m} \lambda_{j m} \ ,
\end{equation}
in terms of which the operators $\hat A_\mu$ become
\begin{equation}
  \label{eq:x6}
  \hat A_\mu\longrightarrow
\Phi_\mu=\sum_{m_\mu=1/2}^{j_\mu} \varphi_{j_\mu m_\mu}
\end{equation}
and the Hamiltonian (better, the normal kernel of)
\begin{equation}
  \label{y7}
  H= \sum_{\nu=1}^L e_\nu \sum_{m_\nu=-j_\nu}^{j_\nu}
\lambda^*_{j_\nu m_\nu} \lambda_{j_\nu m_\nu} -
G \sum_{\mu,\nu=1}^L \Phi^*_\mu\Phi_\nu \ .
\end{equation}

The index of nil-potency of the collective Grassmann variable
$\Phi_\mu$ (to be referred to as $s$-quasi-boson) is $\Omega_\mu$, i.e.,
\begin{equation}
  (\Phi_\mu)^n=0 \ \ \ \mbox{for}\ n>\Omega_\mu \ .
\label{nilpotency}
\end{equation}

We now search for eigenstates of $n$ pairs of fermions 
in the $s$-quasibosons subspace as products of $n$ factors, namely
\begin{equation}
  \label{eq:1}
\psi_n(\Phi^*) = \prod_{k=1}^n {\cal B}^*_k
\end{equation}
where
\begin{equation}
  {\cal B}^*_k = \sum_{\nu=1}^L \beta_\nu^{(k)} \Phi^*_\nu \ ,
\label{psin}
\end{equation}
is a superposition of $s$-quasibosons placed in all the
available levels.

For this scope it is convenient to start from the effective Hamiltonian
\begin{equation}
  \label{eq:y8}
  {\cal H}_{\rm eff}(\varphi^*,\varphi) =
\sum_{\nu=1}^L 2 e_\nu \sum_{m_\nu=1/2}^{j_\nu}
\varphi^*_{j_\nu m_\nu} \varphi_{j_\nu m_\nu} -
G \sum_{\mu,\nu=1}^L \Phi^*_\mu\Phi_\nu \ ,
\end{equation}
coincident with (\ref{y7}) in the $s$-quasibosons subspace spanned by 
the states (\ref{eq:1}).
Indeed while terms like $\lambda^*\lambda$ count the number of particles, 
$\varphi^*\varphi\equiv\lambda^*\lambda^*\lambda\lambda$ counts the number 
of pairs.
The eigenvalue equation then reads
\begin{align}
\nonumber
H \psi_n(\Phi^*) &=
\int \left[d\lambda' d{\lambda^*}^\prime\right]
{\cal H}_{\rm eff}(\varphi^*,\varphi^\prime)
\exp\left(\sum_{\mu=1}^L \sum\limits_{m_\mu=-j_\mu}^{j_\mu}
\lambda^*_{j_\mu m_\mu}\lambda_{j_\mu m_\mu}^\prime\right)\\
&\times 
\exp\left(-\sum_{\mu=1}^L  \sum\limits_{m_\mu=-j_\mu}^{j_\mu}
{\lambda^*}^\prime_{j_\mu m_\mu}\lambda_{j_\mu m_\mu}^\prime\right)
\psi_n({\Phi^*}^\prime)
= E_n  \psi_n(\Phi^*)\ ,
\label{eq:12}
\end{align} 
where
\begin{equation}
  [d{\lambda^*}^\prime d\lambda^\prime]
\equiv \prod_{\nu=1}^L \prod_{m_\nu=-j_\nu}^{j_\nu}
d{\lambda^*}^\prime_{j_\nu,m_\nu} d\lambda^\prime_{j_\nu,m_\nu}\ .
\end{equation}
By expanding the exponentials in \eqref{eq:12}, 
only the even powers, hence only
the $\varphi$ variables, survive. Thus \eqref{eq:12} can be rewritten as
\begin{equation}
  \label{eq:13}
\int [d{\varphi^*}^\prime d\varphi^\prime]{\cal H}_{\rm eff}
(\varphi^*,\varphi^\prime)
{\cal M}(\varphi^*+{\varphi^*}^\prime,\varphi^\prime) \psi_n({\Phi^*}^\prime
) = E_n  \psi_n(\Phi^*)\ ,
\end{equation}
where
\begin{equation}
  {\cal M}(\varphi^*,\varphi)  \equiv
\exp\left(\sum_{\mu=1}^L\sum_{m_\mu=1/2}^{j_\mu}
\varphi^*_{j_\mu m_\mu}\varphi_{j_\mu m_\mu}\right)\ .
\end{equation}
The integrals over the $\varphi$'s,
relevant for dealing with eq.~\eqref{eq:12},
are listed in~\cite{Pal1,Pal2}.
One gets
\begin{equation}
E_n = \sum_{k=1}^n \eta_k \ ,\ \ \ \
\beta_\mu^{(k)} = \frac{1}{2 e_\mu-\eta_k} \ ,
\end{equation}
the $\eta_k$ being the solutions of the non-linear system
\begin{equation}
\sum_{\mu=1}^L \frac{\Omega_\mu}{2 e_\mu-\eta_k}-
\sum_{\substack{l=1\\l\neq k}}^n \frac{1}{\eta_l-\eta_k}=\frac{1}{G} \ .
\label{system}
\end{equation}

In this paper we confine ourselves to the case of a single pair only.
Then \eqref{system} reduces to a single equation and the wave function reads
\begin{equation}
  \label{eq:14}
  \psi_1(\Phi^*) = \sum_{\nu=1}^L \beta_\nu \Phi^*_\nu \ .
\end{equation}
Since the action of $H$ on \eqref{eq:14} is
\begin{eqnarray}
  \label{eq:16}
      \lefteqn{  H\psi_1(\Phi^*) }\nonumber\\
    &&= \sum_{\nu=1}^L 2 e_\nu \sum_{\mu=1}^L \beta_\mu
    \sum_{m_\nu=1/2}^{j_\nu} \varphi^*_{j_\nu m_\nu}
    \int [d{\varphi^*}^\prime d\varphi^\prime]
    \varphi^\prime_{j_\nu m_\nu} 
    {\cal M}(\varphi^*+{\varphi^*}^\prime,\varphi^\prime)
    {\Phi^*}^\prime_{\mu}
\nonumber\\
&&- G\sum_{\mu,\nu=1}^L \sum_{\rho=1}^L \beta_\rho
        \Phi^*_\mu \int [d{\varphi^*}^\prime d\varphi^\prime]
        \Phi^\prime_\nu 
        {\cal M}(\varphi^*+{\varphi^*}^\prime,\varphi^\prime)
        {\Phi^*}^\prime_\rho
\nonumber\\
    &&= \sum_{\nu=1}^L 2 e_\nu \beta_\nu \Phi^*_\nu
 -G\sum_{\mu,\nu=1}^L  \Omega_\mu\beta_\mu \Phi^*_\nu\ ,
\end{eqnarray}
the Schr\"odinger equation becomes
\begin{equation}
  \label{eq:18}
  \sum_{\nu=1}^L \left[
    (2 e_\nu - E) \beta_\nu - G \sum_{\mu=1}^L  \Omega_\mu\beta_\mu
  \right] \Phi^*_\nu = 0
\end{equation}
which implies
\begin{equation}
  \label{eq:19}
  (2 e_\nu - E) \beta_\nu - G \sum_{\mu=1}^L  \Omega_\mu\beta_\mu = 0
\ \ \ \ \ \forall\nu=1,L\ .
\end{equation}
Since $\sum_{\mu=1}^L  \Omega_\mu\beta_\mu $ does not depend on the
index $\nu$, it follows that the coefficients $\beta_\nu$ are 
\begin{equation}
  \label{eq:21}
  \beta_\nu=\frac{C}{2 e_\nu - E}\ ,
\end{equation}
$C$ being a normalisation factor.
Inserting  \eqref{eq:21} into (\ref{eq:19}) we then get
 the well-known eigenvalue equation (referred to in the following as
secular equation)~\cite{Ro-70-B}
\begin{equation}
  \label{eq:20}
   \sum_{\nu=1}^L \frac{\Omega_\nu}{2 e_\nu - E} = \frac{1}{G}\ ,
\end{equation}
which yields $L$ eigenvalues $E^{(\mu)}$
($1\leq \mu\leq L$), the corresponding components  $\beta_\nu^{(\mu)}$
of the wave function being given by eq.~\eqref{eq:21}.

Actually the normalisation of the state
\begin{equation}
\left(\psi_1^*,\psi_1\right) = \sum_{\nu,\nu'=1}^L \beta^*_\nu\beta_{\nu'}
\left(\Phi_\nu,\Phi^*_{\nu'}\right)=\sum_{\nu=1}^L \Omega_\nu |\beta_\nu|^2=1
\end{equation}
suggests to introduce the coefficients
\begin{equation}
\tilde \beta_{\nu}^{(\mu)} = \sqrt{\Omega_{\nu}} \beta_{\nu}^{(\mu)}\ .
\end{equation}

\begin{figure}[H]
  \begin{center}
    \epsfig{file=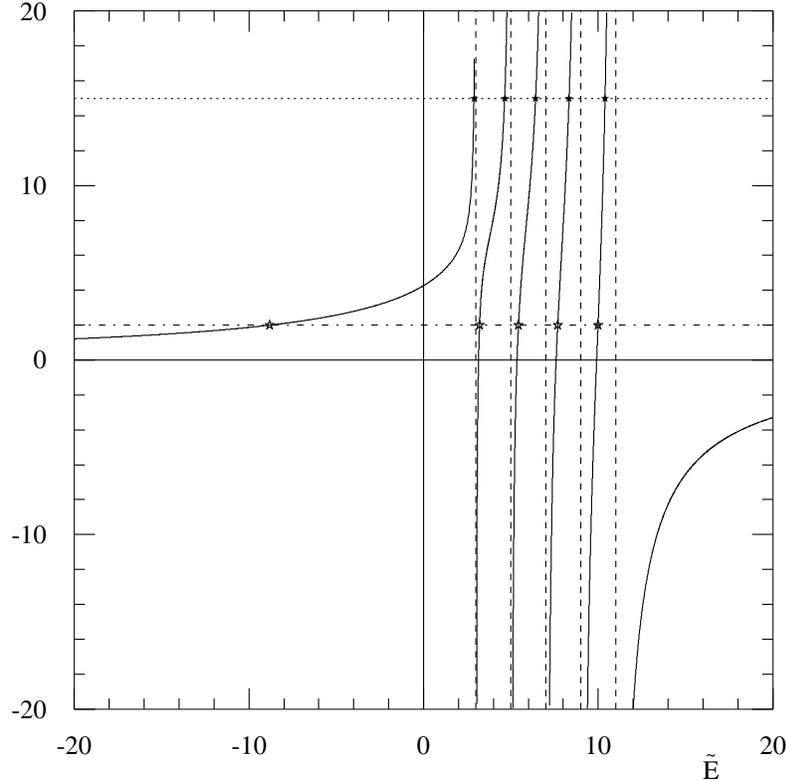,width=13cm}
    \caption{Graphical solution of the eq.~\protect\eqref{eq:20}
      for the case of a harmonic oscillator well with 5 levels.
      The values of $\tilde G$  are 1/2 (dot-dashed) 
      and 1/15 (dotted). All quantities are in dimensionless units
      ($\tilde E=E/\hbar\omega_0$).}
    \label{fig:1}
  \end{center}
\end{figure}

It is straightforward to solve equation \eqref{eq:20} numerically: 
the solutions can be graphically displayed as the intersections
of the lhs of \eqref{eq:20} with the straight line $E=1/G$.
In Fig.~\ref{fig:1} this is done for 5 levels of a 3-dimensional harmonic 
oscillator and for two typical values of 
$\tilde G\equiv G/\hbar\omega_0$.
In the figure two classes of states (labelled as $k=0,\dots,{\cal N}-1$)
appear: the first one embodying the $k=0$ state, which lies below the lowest
single particle level for an attractive interaction and corresponds to a
collective state;
the other embodies the so-called ``trapped'' solutions $(1\le k \le 4)$, 
which lie in between the single particle levels.
\begin{figure}[ht]
  \begin{center}
    \epsfig{file=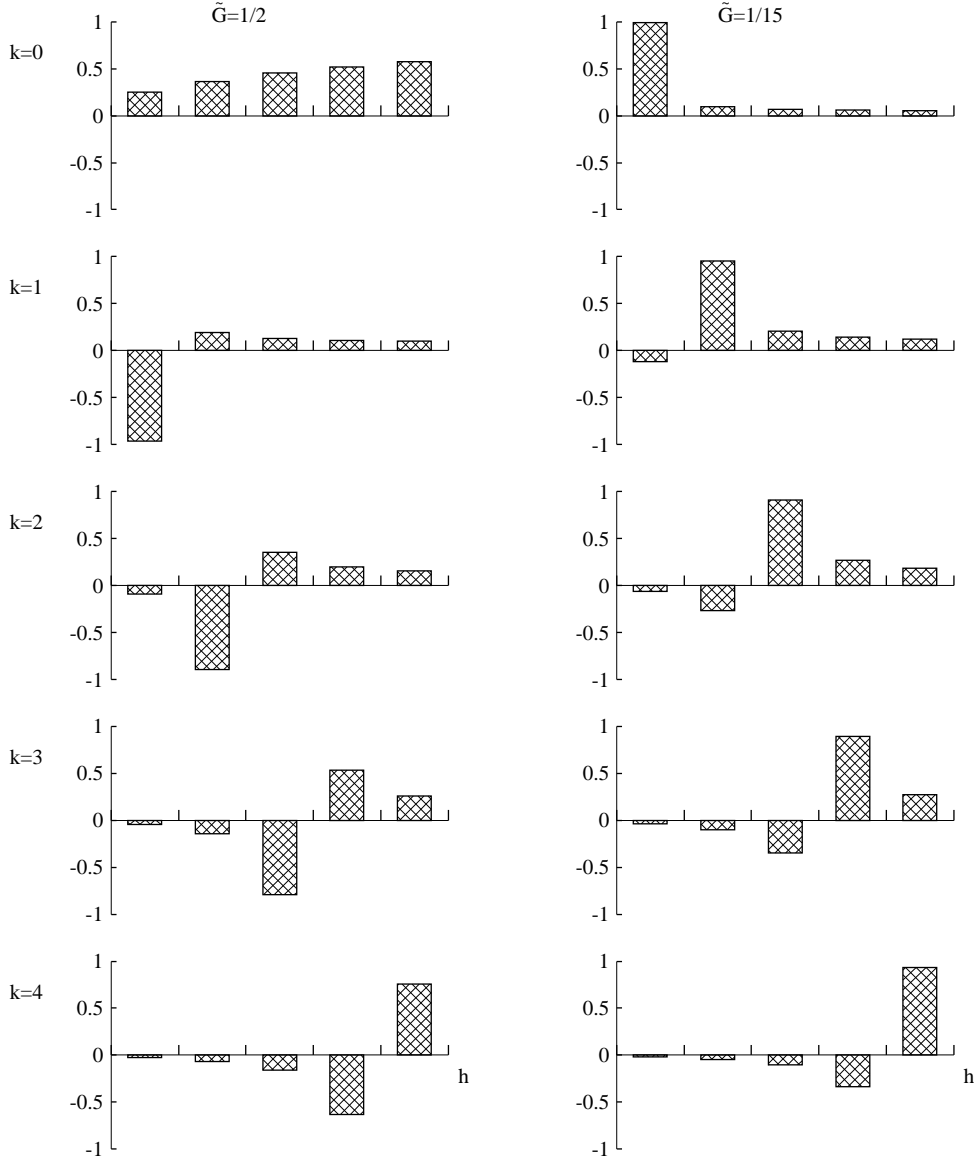,height=16cm,width=13cm}
    \caption{The wave functions components, $\tilde\beta_h^{(k)}$,
      of the collective and trapped states (from
      top to bottom) for $\tilde G=1/2$ (left panel) and $\tilde G=1/15$
      (right panel)
      in the same case as in Fig. \protect\ref{fig:1}.}
    \label{fig:2}
  \end{center}
\end{figure}
\afterpage{\clearpage}

The collective and single particle
character of the eigenstates is apparent in the histograms of
Fig.~\ref{fig:2}, where the coefficients $\tilde \beta_h^{(k)}$
are displayed.
The states on the left panel correspond to the coupling constant
$\tilde G=1/2$, those on the right panel to $\tilde G=1/15$.
From the top
to the bottom the coefficients of the collective ($k=0$)
and of the trapped  ($ 1\le k\le 4$) states are shown.
The bars, from left to right, yield the components $h=0,\cdots 4$ of the
wave functions
\begin{equation}
\psi_1^{(k)}(\Phi^*)=\sum_{h=0}^{{\cal N}-1}
\frac{\tilde\beta_h^{(k)}}{\sqrt{\Omega_h}}\Phi^*_h\ .
\end{equation}

In the $\tilde G=1/2$ case all the $\tilde \beta_h^{(0)}$ are sizable,
reflecting the high degree of collectivity of the state, although the weight of
the $s$-quasibosons living in 
single particle states with higher degeneracy, and hence more distant from the
collective state, tends to be greater.
For $\tilde G=1/15$ the force is too weak to organise any collective motion.

Concerning the eigenstates of the trapped solutions when
$\tilde G$ is large, the $k$=1 and 2 states approach the lower unperturbed
levels, and so do their wave functions: in fact in the left panel 
one sees that the dominant component of the latter corresponds to $h=k-1$.
Instead the energies of the $k$=3 and 4 states occur close to the middle of 
two unperturbed energies: hence their wave functions display
{\em two} dominant components.
For small $\tilde G$ the energies of the trapped solutions
remain close to the unperturbed energies, hence
their wave function almost coincide with the unperturbed $h=k$ state.

\section{The collective solution}
\label{sec:coll}

We first consider the collective solution in
the extreme situation where all the levels coalesce. Then
eq.~\eqref{eq:20} reduces to
\begin{equation}
  \label{eq:6.1a}
  \frac{\Omega}{2\overline e-E}=\frac{1}{G}
\end{equation}
(with $\Omega=\sum_\nu\Omega_\nu$ and $e_\nu=\overline e\ \forall \nu=1,
\dots,L$), entailing
\begin{equation}
  \label{eq:6.1b}
  E=2\overline e-\Omega G\;.
\end{equation}
Thus in this limit only the collective state survives.
Clearly \eqref{eq:6.1b} remains a good approximation when the degeneracies 
of the $L$ levels are lifted only if their spread in energy is small with 
respect to $\Omega G$.
An accurate analytic solution when the distribution of the levels
is arbitrary will be derived in the next subsection in both the
large and small $G$ limits.

\subsection{The `strong coupling' limit}
\label{sec:coll_strong}

For large $G$ it is convenient to 
recast eq.~\eqref{eq:20} as 
\begin{equation}
\label{eq:6.1bis}
\begin{split}
\sum_\nu\frac{\Omega_\nu}{2e_\nu-E}& =
-\frac{1}{E-2\bar e} \sum_\nu \frac{\Omega_\nu}{1-2
\dfrac{e_\nu-\bar e}{E-2\bar e}}=\frac{1}{G}\;.
\end{split}
\end{equation}
and expand in the
parameter $2(e_\nu-\bar e)/(E-2\bar e)$.
Defining
\begin{equation}
\bar e=\frac{\sum_\nu \Omega_\nu e_\nu}{\Omega}
\label{ebar}
\end{equation}
in leading order we get
\begin{equation}
E_0=2\bar e-\Omega G\ ,
\end{equation}
which coincides with the degenerate case value \eqref{eq:6.1b}.
Importantly, owing to the definition \eqref{ebar}, the next-to-leading order 
correction vanishes.

To proceed further we rewrite \eqref{eq:6.1bis} as 
\begin{equation}
  \label{eq:n6.3}
  -\frac{\Omega}{E-2\bar e}\sum_{n=0}^\infty
  \frac{2^n M^{(n)}}{(E-2\bar e)^n}=
-\frac{\Omega}{E-2\bar e}\sum_{n=0}^\infty
\left(\frac{G\Omega}{E-2\bar e}\right)^n m^{(n)} \alpha^n=
\frac{1}{G}\;,
\end{equation}
where the generalised moments of the distribution of
single particle levels $e_\nu$, namely
\begin{equation}
  \label{eq:n6.2}
M^{(n)}=\sigma^n m^{(n)}=\frac{1}{\Omega}\sum_\nu\Omega_\nu(e_\nu-\bar e)^n\ ,
\end{equation}
have been introduced together with the expansion parameter
\begin{equation}
  \label{eq:n6.7}
  \alpha=\frac{2\sigma}{G\Omega}
\end{equation}
and the variance
\begin{equation}
  \label{eq:n6.2a}
\sigma =\sqrt{\frac{1}{\Omega}\sum_\nu\Omega_\nu(e_\nu-\bar e)^2}
\end{equation}
of the level distribution.
The `strong coupling' regime then corresponds to $\alpha\ll 1$.
Note that the second moment is the square of the variance $\sigma$, 
whereas the third, $M^{(3)}=\sigma^3\gamma$, and fourth,
$M^{(4)}= (c+3) \sigma^4$, moments are related to the skewness $\gamma$
and to the kurtosis $c$ of the distribution, respectively.

In a perturbative scheme, setting $E^{(n)} = E^{(n-1)} +\delta$
and linearising in $\delta$, we get for the collective energy
\begin{equation}
  \label{eq:n6.18}
  \frac{E-2\bar e}{G\Omega}=-1-\alpha^2+\gamma\alpha^3-(1+c)\alpha^4 +
{\cal O}(\alpha^5)\ ,
\end{equation}
or, equivalently,
\begin{equation}
  \label{eq:n6.17}
  E\simeq -G\Omega+2\bar e-\frac{4\sigma^2}{G\Omega}+\frac{8\gamma\sigma^3}
  {G^2\Omega^2}-\frac{16(1+c)\sigma^4}{G^3\Omega^3}\ ,
\end{equation}
an expression valid for $\alpha\ll 1$ and independent of the single particle 
energies distribution.

\subsection{The `weak coupling' limit}
\label{sec:coll_weak}

When $\alpha > 1$
the collectivity is very weak (see Fig.~\ref{fig:2}, right panel)
and in \eqref{eq:6.1bis} the term $\nu=1$ dominates the sum. 
Thus by separating the latter we rewrite \eqref{eq:6.1bis} in the form
\begin{equation}
\sum_{\nu=2}^{L} \frac{\Omega_\nu}{2(e_\nu -e_1)-(E-2e_1)}+
\frac{\Omega_1}{2e_1-E} =\frac{1}{G}\;.
\label{eq:100}
\end{equation}
Now, for $G\simeq 0$, the quantity $2e_1-E$ is of the order of $G$:
hence the last term in the lhs of \eqref{eq:100} dominates.
Thus the leading contribution to the energy is
\begin{equation}
E_0=2e_1-G\Omega_1\ .
\end{equation}
Factorising then $2(e_\nu -e_1) $ in the denominator of \eqref{eq:100},
expanding in powers of $(E-2e_1)/2(e_\nu-e_1)$ and
introducing the inverse generalised moments
\begin{equation}
M^{(-r)}=\frac{m^{(-r)}}{\sigma^r}=
\frac{1}{\Omega-\Omega_1}\sum_{\nu=2}^{L}\frac{\Omega_\nu}{(e_\nu-e_1)^r}
\end{equation}
we get
\begin{multline}
\frac{G(\Omega-\Omega_1)}{2\sigma} \left[
m^{(-1)}+ \frac{E-2e_1}{2\sigma} m^{(-2)}+\left(\frac{E-2e_1}{2\sigma}\right)^2
m^{(-3)}+\cdots \right]\\ - \frac{G\Omega_1}{E-2e_1}=1\ .
\end{multline}
Proceeding then as in section~\ref{sec:coll_strong},
we next expand in powers of $\beta\equiv m^{(-1)}/\alpha'$,
being $1/\alpha' =G(\Omega-\Omega_1)/2\sigma$.
At the order $\beta^3$ we obtain
\begin{eqnarray}
\label{eq:a47}
\frac{E-2e_1}{G\Omega_1} &\simeq&-\left\{
1+\beta + \beta^2\left[
1-\frac{m^{(-2)}}{(m^{(-1)})^2}\frac{\Omega_1}{\Omega-\Omega_1}\right]
\right.\\
& +& \left. \beta^3\left[
1-3\frac{m^{(-2)}}{(m^{(-1)})^2}\frac{\Omega_1}{\Omega-\Omega_1}
+\frac{m^{(-3)}}{(m^{(-1)})^3}\left(\frac{\Omega_1}{\Omega-\Omega_1}\right)^2
\right]
\right\}\nonumber\;,
\end{eqnarray}
where $\alpha'$ does not coincide with $1/\alpha$ because 
the lowest unperturbed state has been separated out.
Note that not only where the `strong coupling' expansion fails
the `weak coupling' one holds valid, but also an overlap
region appears to exist where both expansions yield quite accurate results,
being at the same time $\alpha <1$ and $\beta<1$.

\subsection{The Euler-McLaurin approximation}
\label{sec:coll_EML}

An interesting approach to the pairing problem is offered by
the  Euler-McLaurin formula~\cite{SaGe-60-B}. 
It enables us to replace the sum with an integral, i.e.\footnote{The function
$R$ was first fixed by Euler to be $\int_a^b (u-[u]-1/2) f'(u) du$.
McLaurin provided more accurate expressions for it.}
\begin{equation}
\sum_{\nu=a}^{b} f(\nu)= \frac{1}{2} f(a) + \frac{1}{2} f(b) +
\int_{a}^b f(u) du + R \ .
\label{maclaurin}
\end{equation}
The formula \eqref{maclaurin} holds valid if $f(\nu)$, $\nu$ to be viewed as
a complex variable, is analytic in the strip $a\leq \mbox{Re}\nu \leq b$ ($a$ 
and $b$ being integers).
Neglecting the last term in the rhs,
we shall illustrate the use of \eqref{maclaurin} in the case of
the harmonic oscillator well, whose
levels we label with an index $k=0,1,\cdots {\cal N}-1$.
For this potential, from the well-known unperturbed energies
\begin{equation}
e_k\equiv \tilde e_k \hbar \omega_0 = (k+ 3/2) \hbar \omega_0
\label{ekHO}
\end{equation}
and associated pair degeneracies
\begin{equation}
\Omega_k = (k+1)(k+2)/2\ ,
\label{OmkHO}
\end{equation}
the total degeneracy $\Omega$, the average energy $\bar e$ and the variance 
$\sigma$ are found to be
\begin{equation}
  \label{eq:3.4.4}
  \Omega=\frac{1}{6}{\cal N}({\cal N}+1)({\cal N}+2)\ ,
\end{equation}
\begin{equation}
  \label{eq:3.4.3}
  \overline e=\frac{3}{4}({\cal N}+1)\hbar \omega_0
\end{equation}
and
\begin{equation}
\sigma^2 = \frac{3}{80}({\cal N}-1)({\cal N}+3) (\hbar \omega_0)^2\ ,
\end{equation}
respectively.
The above entail
\begin{equation}
\alpha^{(h.o.)}=
\frac{3\hbar\omega_0}{G} \sqrt{\frac{3}{5}}
\frac{\sqrt{({\cal N}-1)({\cal N}+3)}}{{\cal N}({\cal N}+1)({\cal N}+2)}\ .
\label{eq:alpha_ho}
\end{equation}
Incidentally, from \eqref{eq:alpha_ho} the coincidence of 
\eqref{eq:n6.17} at the order
$\alpha^2$ with the finding of ref.~\cite{BaFoMoQu-01} follows.

Before exploiting \eqref{maclaurin}
we recast \eqref{eq:20} using the digamma functions and 
the dimensionless variables
$\tilde G\equiv G/\hbar \omega_0$ and $\tilde E\equiv E/\hbar \omega_0$; we get
\begin{equation}
\label{Roweho}
\begin{split}
 F^{{\cal N}}(\tilde E)&\equiv
 \sum\limits_{N=0}^{{\cal N}-1} \frac{(N+1)(N+2)}{2N+3-\tilde E}
\\
&=\frac{{\cal N}(2+\tilde E+{\cal N})}{4}
  +\frac{\tilde E^2-1}{8}
  \left[\psi\left({\cal N}+\frac{3-\tilde E}{2}\right)
    -\psi\left(\frac{3-\tilde E}{2}\right)
    \right]
\\
&=\frac{{\cal N}(2+\tilde E+{\cal N})}{4}\\
  &+\frac{\tilde E^2-1}{8}
  \left[\psi\left({\cal N}+\frac{1-\tilde E}{2}\right)-
  \psi\left(\frac{3-\tilde E}{2}\right)
    +\frac{2}{1+2{\cal N}-\tilde E}\right]
 =\frac{2}{\tilde G}\ .
  \end{split}
\end{equation}

In the `strong coupling' limit the collective solution is strongly
pushed down in 
energy, hence the use of the asymptotic formula
\begin{equation}
  \label{eq:3.16}
  \psi(z) \sim \ln z -\frac{1}{2z} - \sum_{k=1}^\infty 
\frac{B_{2k}}{2kz^{2k}}\ ,
\end{equation}
the $B_{2k}$ being the Bernoulli numbers, is appropriate.
It is then remarkable that by inserting the $\psi$ as given by the first
two terms on the rhs of \eqref{eq:3.16} into \eqref{Roweho} one 
obtains the same expression provided by the Euler-McLaurin formula for the 
harmonic oscillator, namely
\begin{equation}
  \begin{split}
    F^{{\cal N}}_{\rm E-McL}(\tilde E)&=
    \frac{{\cal N}(2+\tilde E+{\cal N})}{4}
    \\&+ \frac{\tilde E^2-1}{8}
    \left[\frac{1}{3-\tilde E} + \frac{1}{2{\cal N}+1-\tilde E} +
      \ln{\frac{2{\cal N}+1-\tilde E}{3-\tilde E}}\right]\ .
    \end{split}
\label{nocarciofo}
\end{equation}
In the tables \ref{tab:5} and \ref{tab:6} we refer to the eigenvalues  
obtained using \eqref{nocarciofo} as $ML$.
Furthermore, being the term $z^{-3}$ absent in \eqref{eq:3.16}, 
to keep in the expansion \eqref{eq:3.16} also the $k=1$ terms yields
excellent results.
We refer to this approximation as $ML2$.

\subsection{Numerical results for the collective energy}
\label{sec:coll_results}

In this subsection we present our predictions for the collective energy,
essentially in the harmonic oscillator case, for
$\tilde G$=0.1 and 0.2 (a realistic estimate for atomic nuclei
with mass number $A>100$).
Our results, corresponding to eqs.~\eqref{eq:n6.18}, \eqref{eq:a47} and
to the solution eq.~\eqref{Roweho} with $F^{\cal N}$ given by 
\eqref{nocarciofo},
are shown in tables~\ref{tab:5}, \ref{tab:6}
and compared with the exact solutions
for some values of ${\cal N}$.

\begin{table}[H]
\begin{center}
\begin{tabular}{||c|c|c|c|c|c|c|c|c|c||}
\hline\hline  \raisebox{0pt}[14pt]{} \raisebox{-7pt}[16pt]{} 
${\cal N}$ & $\tilde E_{\rm exact}$ &
${\cal O}(\alpha^2)$
&${\cal O}(\alpha^4)$ & $ \alpha$
&${\cal O}(\beta)$ & ${\cal O}(\beta^3)$ 
& $\beta $
& $\tilde E_{\rm ML}$ & $\tilde E_{\rm ML2} $\\
\hline\hline
   \footnotesize 2 & \footnotesize 2.883 & --  & -- &\footnotesize 2.17
& \footnotesize2.885 & \footnotesize2.883& \footnotesize0.15
& \footnotesize2.927  & \footnotesize 2.822 \\
\footnotesize  3 & \footnotesize 2.860  & --& -- &\footnotesize 1.34
& \footnotesize2.87 & \footnotesize2.860 &\footnotesize0.3
& \footnotesize2.909   &    \footnotesize 2.800\\
\footnotesize   4 & \footnotesize 2.820 & \footnotesize 3.925
& \footnotesize 1.479 &\footnotesize 0.89 & \footnotesize 2.853
& \footnotesize 2.825 & \footnotesize 0.47
&\footnotesize2.877  &\footnotesize  2.763 \\
\footnotesize   6 & \footnotesize 2.529 & \footnotesize 3.695
& \footnotesize 2.930 &\footnotesize 0.46 & \footnotesize 2.814
&\footnotesize 2.681&\footnotesize0.86
&\footnotesize 2.590  &  \footnotesize 2.500\\
 \footnotesize  8 & \footnotesize0.1361  & \footnotesize 0.5375
& \footnotesize 0.2161 &\footnotesize 0.28
& -- &-- & \footnotesize1.35
& \footnotesize 0.1338   &  \footnotesize 0.1362\\
\footnotesize   10 & \footnotesize-6.477  & \footnotesize -6.298
& \footnotesize -6.461 &\footnotesize 0.19
& --&-- &\footnotesize1.95
& \footnotesize-6.459    & \footnotesize -6.477\\
 \footnotesize  12 & \footnotesize-17.68 & \footnotesize -17.58
&  \footnotesize-17.68 &\footnotesize 0.14
& --&-- &\footnotesize 2.63 &\footnotesize-17.64
&\footnotesize  -17.68
\\ \hline\hline
\end{tabular}
\caption{Comparison between the exact and the approximate collective energy
for the harmonic oscillator case, for various values of ${\cal N}$ and 
${\tilde G}=0.1$. The energies are in units of $\hbar\omega_0$.
Columns 3 and 4: strong coupling; columns 6 and 7: weak coupling; columns
9 and 10: Euler-McLaurin approximations.}
\label{tab:5}
\end{center}
\end{table}
\begin{table}[H]
\begin{center}
\begin{tabular}{||c|c|c|c|c|c|c|c|c|c||}
\hline\hline\raisebox{0pt}[14pt]{} \raisebox{-7pt}[16pt]{} 
${\cal N}$ & $\tilde E_{\rm exact}$ &${\cal O}(\alpha^2)$
&${\cal O}(\alpha^4)$ & $ \alpha$
&${\cal O}(\beta)$ & ${\cal O}(\beta^3)$ & $\beta $
& $\tilde E_{\rm ML}$ & $\tilde E_{\rm ML2} $\\
\hline\hline
 \footnotesize  2 & \footnotesize2.728 & --& -- & \footnotesize1.1
& \footnotesize 2.74 & \footnotesize 2.727 & \footnotesize0.3
& \footnotesize2.796   &  \footnotesize  2.677\\
\footnotesize   3 & \footnotesize 2.583 & \footnotesize 3.1
&\footnotesize 2.185 & \footnotesize0.67
& \footnotesize 2.68 &\footnotesize2.589 & \footnotesize0.6&
\footnotesize2.653    & \footnotesize  2.547\\
\footnotesize  4  & \footnotesize2.182   & \footnotesize2.713
&\footnotesize2.243  & \footnotesize0.44
& \footnotesize 2.613 &\footnotesize 2.318 &\footnotesize 0.93
&\footnotesize 2.224 &\footnotesize 2.171\\
 \footnotesize  6 & \footnotesize-1.480 & \footnotesize -1.303
& \footnotesize-1.461  & \footnotesize0.23
&--  &-- &\footnotesize 1.73
&\footnotesize-1.478  &\footnotesize-1.480\\
 \footnotesize  8 & \footnotesize-11.06 & \footnotesize -10.98
& \footnotesize -11.05  & \footnotesize0.14
& --&-- &\footnotesize2.71
& \footnotesize -11.01 &\footnotesize -11.06\\
 \footnotesize 10 &\footnotesize -27.94 & \footnotesize -27.90
& \footnotesize -27.94  &\footnotesize 0.10
&-- &--&\footnotesize 3.88
&\footnotesize-27.85 & \footnotesize -27.94\\
 \footnotesize 12 & \footnotesize-53.66 & \footnotesize -53.64
& \footnotesize -53.66  &\footnotesize 0.07
&--  & --& \footnotesize5.25
& \footnotesize-53.53 & \footnotesize -53.66 \\ \hline\hline
\end{tabular}
\caption{Same as table \protect\ref{tab:5} for  ${\tilde G}=0.2$.}
\label{tab:6}
\end{center}
\end{table}
\begin{figure}[H]
  \begin{center}
    \epsfig{figure=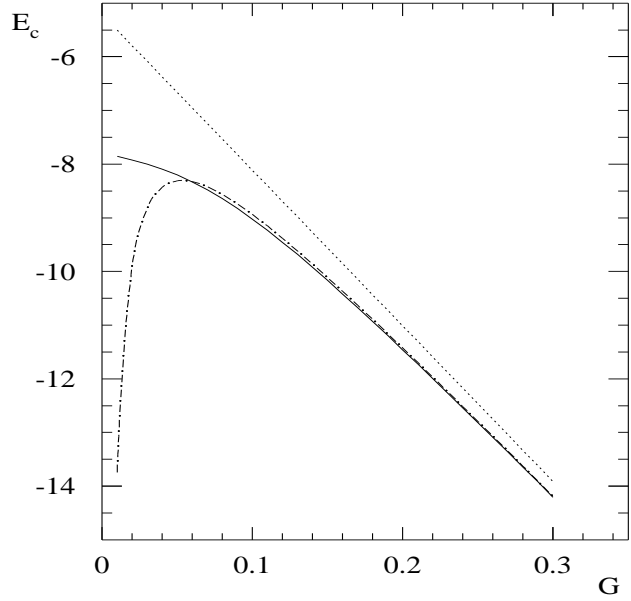,width=10cm,height=10cm}
    \caption{The collective energy of $^{210}Pb$ as a function of the pairing
      strength $G$ (in MeV). Solid line: exact numerical
      solution; dotted: degenerate solution (order $\alpha^0$);
      dot-dashed: the approximation (\protect{\ref{eq:n6.18}}) at the order 
      $\alpha^2$.}
    \label{fig:3}
    \end{center}
\end{figure}

The approximate energies \eqref{eq:n6.18} are
in good agreement with the exact ones for $\tilde G\ge 0.2$.
Instead for smaller $\tilde G$ and $4\le{\cal N}\le 8$ the collective energy
in the `weak coupling' limit, namely eq.~\eqref{eq:a47},
is in better touch with the exact values.
The existence of a region where both the `strong' and the `weak' 
approximations hold valid, is apparent.
We also quote in the tables \ref{tab:5}, \ref{tab:6}
the results obtained with the McLaurin approximation in first and second
order, as previously discussed.

To address a situation of physical interest,
we consider the case of $^{210}Pb$, where one pair of neutrons lives in the
$N=6$ shell, set up with the s.p. levels $d_{3/2}$, $g_{7/2}$, $s_{1/2}$, 
$d_{5/2}$, $j_{15/2}$, $i_{11/2}$ and $g_{9/2}$ of energies --1.5, --1.6,
--2.0, --2.3, --2.4, --3.1 and --3.9 MeV, respectively.
The total degeneracy $\Omega$,
the average energy value $\bar e$, see \eqref{ebar}, and the variance $\sigma$
then turn out to be 29, --2.61 and 0.77 MeV, respectively.
Since for lead $ G\simeq 0.1$ MeV we obtain $\alpha=0.53 $ and $\beta= 0.89$.
Hence we are in the `strong coupling' limit.

In Fig.~\ref{fig:3} we compare the collective energy \eqref{eq:n6.18}
for  $^{210}Pb$,
at order $\alpha^0 $ and at order $\alpha^2$,
with the exact numerical solution as functions of $G$:
we see that, in the physical range of $G$, an excellent accord with
the exact result is obtained when the term $\alpha^2$ is included.

\section{Trapped solutions}
\label{sec:trapped}

\subsection{The general case}
\label{sec:4.1}

The evaluation of the trapped eigenvalues can be easily performed 
numerically. This approach however
hides the interesting pattern the trapped solutions display, 
already suggested in~\cite{BaFoMoQu-01} where they were associated with a
quantum number and a parabolic behaviour in the latter
was shown to occur for a harmonic oscillator well. 

To shed light on this aspect of the pairing problem we first observe that,
irrespectively of the strength of the interaction, 
 $2e_{\nu-1}<E^{(\nu)}<2e_\nu$, $E^{(\nu)}$ being the energies of the trapped
solutions ($\nu=2,\dots L$).

Next we define a new variable $z^{(\nu)}$ according to
\begin{equation}
  \label{eq:4.01}
  E^{(\nu)}=2e_{\nu-1}+2 z^{(\nu)}(e_\nu-e_{\nu-1})
\end{equation}
(clearly $z^{(\nu)}\in(0,1)$) and 
select from the secular equation the terms associated with the poles in
$2e_{\nu-1}$ and $2e_\nu$, thus getting
\begin{equation}
  \label{eq:4.02}
  \frac{\Omega_{\nu-1}}{z^{(\nu)}}+\frac{\Omega_\nu}{z^{(\nu)}-1}=
  \varphi_\nu(z^{(\nu)})
\end{equation}
with 
\begin{eqnarray}
\nonumber
  \varphi_\nu(z^{(\nu)})&=&\sum_{\substack{\mu=1\\ \mu\not=\nu,\nu-1}}
  \frac{\Omega_\mu}{\dfrac{e_\mu-e_{\nu-1}}{e_\nu
      -e_{\nu-1}}-z^{(\nu)}}-\frac{2(e_\nu
      -e_{\nu-1})}{G}
\\
  \label{eq:4.03}
&=&\sum_{\substack{\mu=1\\ \mu\not=\nu,\nu-1}} \zeta(\mu,z)-
\frac{2(e_\nu-e_{\nu-1})}{G} \ .
\end{eqnarray}

We have thus isolated in the lhs of \eqref{eq:4.02}
the poles trapping the solution $  z^{(\nu)}$ in the interval $(0,1)$.
Now we approximate the rhs. To this purpose we expand in powers of
$z^{(\nu)}$ ($\mu\not=\nu,\nu-1$):

\begin{equation}
  \label{eq:4.05}
  \zeta(\mu,z)=
  \begin{cases}
\sum\limits_{n=0}^\infty\dfrac{\Omega_\mu z^n}
  {\left(\dfrac{e_\mu-e_{\nu-1}}{e_\nu
      -e_{\nu-1}}\right)^{n+1}}\ 
\ \ \ \ \mbox{if}\ \ \mu>\nu
\\
\\
\sum\limits_{n=0}^\infty\dfrac{\Omega_\mu (1-z)^n}
  {\left(\dfrac{e_\nu-e_\mu}{e_\nu-
      e_{\nu-1}}\right)^{n+1}}
\ \ \ \ \ \ \mbox{if}\ \ \mu<\nu-1\ .
    \end{cases}
\end{equation}
Of course \eqref{eq:4.05}, 
truncated at the order $m$, provides a
polynomial approximation for the function $\varphi$.

Next we let the discrete variable $\nu$ become continuous
by means of the Euler-McLaurin formula \eqref{maclaurin}. We thus
obtain for $\varphi$ the expression
\begin{equation}
  \label{eq:4.07}
  \begin{split}
  \varphi^{\rm E-McL}(\nu,z)&=\frac{1}{2}\zeta(1,z)
  +\frac{1}{2}\zeta(\nu-2,z)+\int\limits_1^{\nu-2}
  \zeta(\mu,z)d\mu\\
  &+\frac{1}{2}\zeta(\nu+1,z)+\frac{1}{2}\zeta(L,z)
  +\int\limits_{\nu+1}^{L}
  \zeta(\mu,z)d\mu\\
  &-\frac{2[e(\nu)
      -e(\nu-1)]}{G}\ ,
  \end{split}
\end{equation}
which is analytic in the variable $\nu$, if $e_\nu$ and
$\Omega_\nu$ have been analytically extended.

Approximate expressions for $z^{(\nu)}$ then follow
from the continuous version of \eqref{eq:4.02}, namely
\begin{equation}
  \label{eq:4.09}
  \frac{\Omega({\nu-1})}{z{(\nu)}}+\frac{\Omega(\nu)}{z{(\nu)}-1}=
  \varphi^{\rm E-McL}(\nu,z)\ ,
\end{equation}
which implicitly defines $z$ as a function of the complex variable $\nu$.
The simplest approximation corresponds to replace the rhs with
$\varphi^{\rm E-McL}(\nu,0)$. 
One gets
\begin{equation}
  \label{eq:4.010}
  \begin{split}
  z_0(\nu)&=
  \frac{\varphi^{\rm E-McL}(\nu,0)+\Omega(\nu-1)+\Omega(\nu)} 
{2\varphi^{\rm E-McL}(\nu,0)}
  \\&
    -\frac{\sqrt{[\varphi^{\rm E-McL}(\nu,0)+\Omega(\nu-1)+\Omega(\nu)]^2
      -4\Omega(\nu-1)\varphi^{\rm E-McL}(\nu,0)}}
  {2\varphi^{\rm E-McL}(\nu,0)}\ ,
  \end{split}
\end{equation}
which can be further simplified via a parabolic expansion around, e.g., 
the middle point $\nu=L/2$, namely
\begin{equation}
  \label{eq:4.011}
  z_0(\nu)\simeq z_0(L/2)+\left(\nu-\frac{L}{2}\right)
  {z_0}^\prime(L/2)+\frac{1}{2}\left(\nu-\frac{L}{2}\right)^2
  {z_0}^{\prime\prime}(L/2)\ .
\end{equation}
 
To further proceed a specific model for 
$\Omega(\nu)$ and $e(\nu)$ should be chosen, which we shall do
in the next subsection.
Here we display the numerical results for the case of $^{210}Pb$
already discussed in subsec.~\ref{sec:coll_results}.
\begin{table}[ht] 
\begin{center}
  \begin{tabular}{||c||c|c|c||c|c|c||}
    \hline\hline
    $\nu$&$z^{\rm exact}$&$z_0$&$z_1$&
    $E^{\rm exact}$&$E_0$&$E_1$\\
    \hline\hline
    2&0.560&0.621&0.577 &-6.903&-6.807&-6.876 \\
    3&0.555&0.605&0.577 &-5.423&-5.353&-5.392 \\
    4&0.760&0.762&0.760 &-4.648&-4.648&-4.648 \\
    5&0.929&0.934&0.931 &-4.043&-4.040&-4.041 \\
    6&0.730&0.749&0.748 &-3.416&-3.401&-3.402 \\
    7&0.783&0.782&0.783 &-3.043&-3.043&-3.043 \\
     \hline\hline
  \end{tabular}
  \caption{Exact and approximate energies 
    for the trapped levels in the case of $^{210}Pb$. The $E$'s are in MeV, 
    the $z$'s are pure numbers.}
  \label{tab:Pb-trapped}
\end{center}
\end{table}
In tab.~\ref{tab:Pb-trapped} we display both $z_0(\nu)$ as given by
\eqref{eq:4.010}
and $z_1(\nu)$ (namely the $z(\nu)$ obtained by truncating
\eqref{eq:4.05} at first order and then by solving
the corresponding equation by successive linearisations) together 
with the exact numerical solutions. 
The corresponding energies, see \eqref{eq:4.01},
are also reported. 

\subsection{The harmonic oscillator case}
\label{sec:4.2}

In this subsection the
functions $\Omega(\nu)$ and $e(\nu)$ are those of the harmonic oscillator. 
As in sec.~\ref{sec:coll_EML}, we label the solutions
with the index $k$, $1\leq k\leq {\cal N}-1$ (the value $k=0$ corresponds to
the collective solution). 

Using the dimensionless variables $\tilde E$, $\tilde G$ and
$\tilde z=z/\hbar\omega_0$ we rewrite \eqref{eq:4.02} 
in the Euler-McLaurin approximation \eqref{eq:4.07} obtaining
\begin{equation}
  \label{eq:4.020}
  \frac{\Omega(k-1)}{\tilde z{(k)}}+\frac{\Omega(k)}{\tilde z(k)-1}=
  \varphi^{\rm E-McL}(k,\tilde z{(k)})
\end{equation}
with
\begin{equation}
  \label{eq:4.021}
  \begin{split}
  \varphi^{\rm E-McL}(k,\tilde z)&=
  -\frac{2}{\tilde G}-\frac{3}{4}(4k+2\tilde z+3)+\frac{1}{4}
  ({\cal N}-1)({\cal N}+2k+2\tilde z+3)\\
  &+\frac{k(1-k)}{4(\tilde z+1)}-\frac{(k+2)(k+3)}{4(\tilde z-2)}\\
  &-\frac{1}{2(k+\tilde z-1)}-\frac{{\cal N}({\cal N}+1)}
{4(k+\tilde z-{\cal N})}\\
  &+\frac{1}{2}(k+\tilde z)(k+\tilde z+1)\log\left|\frac{(\tilde z+1)
(k+\tilde z-{\cal N})}{(\tilde z-2)(k+\tilde z-1)}
    \right|\ .
  \end{split}
\end{equation}

\begin{figure}[ht]
  \begin{center}
    \epsfig{file=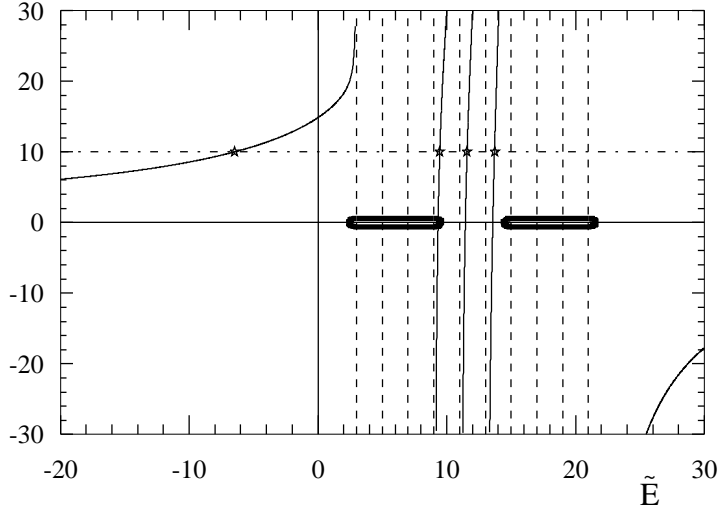,width=13.7cm,height=7.9cm}
    \caption{Graphical solution of the secular equation in the 
      Euler-McLaurin approximation.
      Only the part of the curves outside the cuts (displayed as a black
      area) is drawn.}
    \label{fig:4}
  \end{center}
\end{figure}

The Euler-McLaurin approximation  is not only valid quantitatively,
but also provides the key for studying analytically
limiting cases (like ${\cal N}\to \infty$), which are
helpful in shedding light on the properties
of the trapped eigenvalues. Indeed consider
eq.~\eqref{eq:4.020}: on the lhs it displays the usual two poles in the 
complex plane of $\tilde z$, at 0 and 1, respectively, whereas 
the rhs has poles at $\tilde z$=--1 and 2.
Equivalently, in the variable $\tilde E^{(k)}$, the poles in the range from
$\tilde E^{(k-2)}$ to $\tilde E^{(k+1)}$ are explicitly kept, 
while the others are
simulated by two cuts. The graphical solution is shown in Fig.~\ref{fig:4}
where the quantity 
\begin{equation}
-\frac{\Omega(k-1)}{\tilde z{(k)}}-\frac{\Omega(k)}{\tilde z(k)-1}+
  \varphi^{\rm E-McL}(k,\tilde z{(k)})+\frac{2}{\tilde G}
\end{equation}
(as a function of $\tilde E$) is plotted.
The eigenvalues correspond to the intersections 
with the straight line ${2}/{\tilde G}$.
Four solutions are found: the collective one is of no interest here, 
the other three correspond to the indices $k-1$, $k$ and $k+1$. 
We shall only keep the middle solution, the
most accurate, as it lies far from both the left and the right cut.

\begin{figure}[H]
  \begin{center}
    \epsfig{file=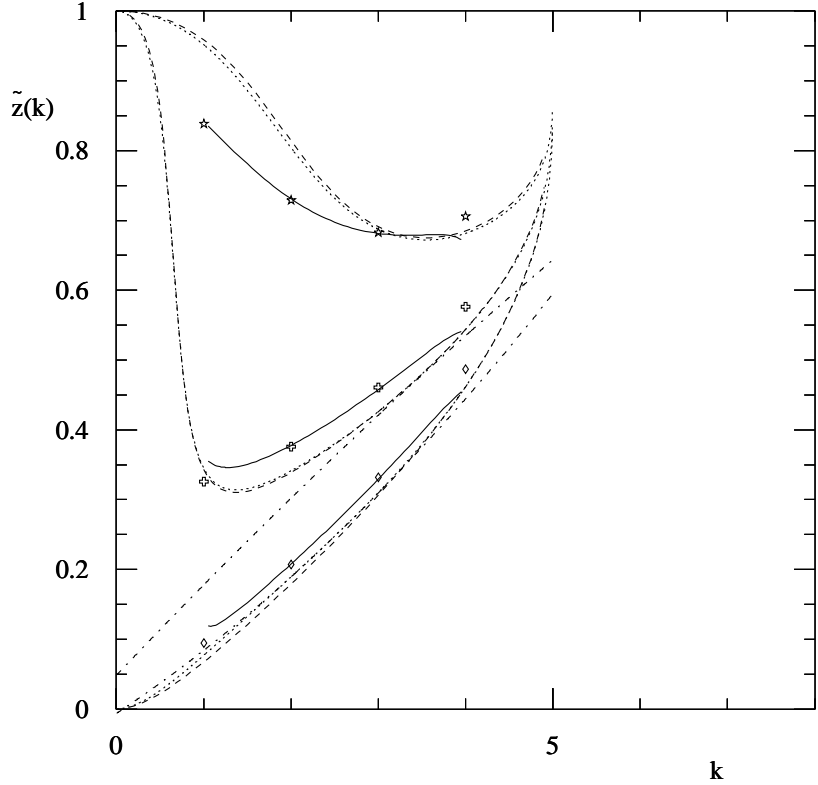,width=13cm,height=13cm}
    \caption{Exact and approximate values for the trapped solutions in the
${\cal N}=5$ case at different values of $\alpha$: diamonds: $\alpha=0.1$, 
crosses: $\alpha=0.4$, 
stars: $\alpha=1$. Solid lines: Euler-McLaurin approximation, 
dashed lines: the same but in the limit ${\cal N}\to\infty$, dotted lines:
approx. eq.~\protect\eqref{eq:4.606}, dash- dotted lines: 
parabolic approximation.}
    \label{fig:6}
  \end{center}
\end{figure}

\begin{figure}[H]
  \begin{center}
    \epsfig{file=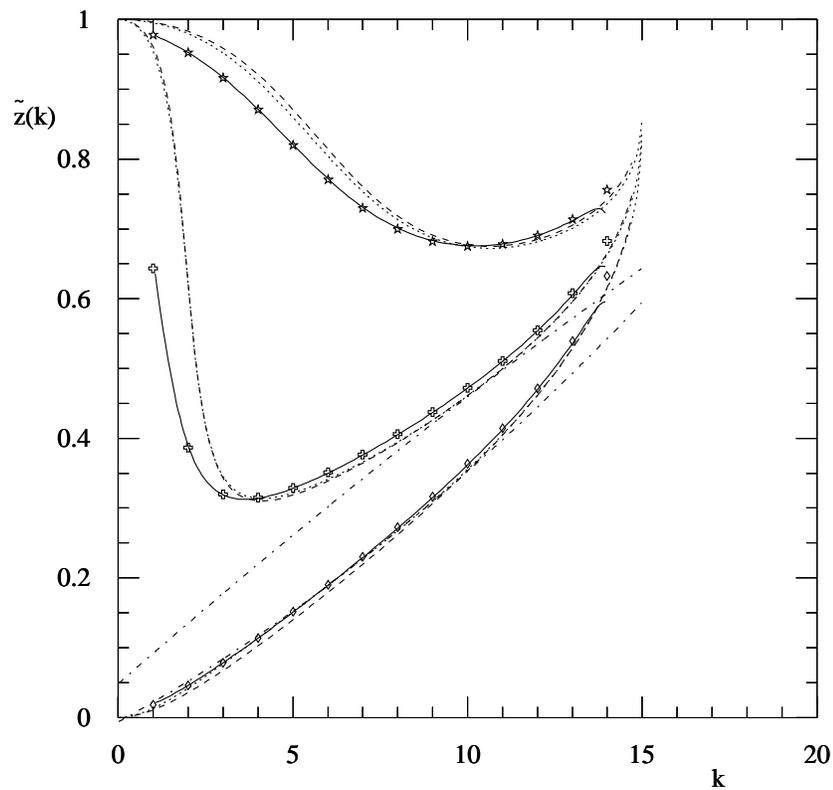,width=13cm,height=13cm}
    \caption{Same as in Fig. \protect\ref{fig:6} but with ${\cal N}=15$}
    \label{fig:7}
  \end{center}
\end{figure}

The results of the Euler-McLaurin approximation for the trapped energies
are shown in Figs.~\ref{fig:6} and \ref{fig:7}, as a solid line. 

We now explore on the basis of
\eqref{eq:4.020} the ${\cal N}\to\infty$ limit. For this purpose 
we introduce the variable
\begin{equation}
  \label{eq:9.201}
  \lambda=\frac{k}{{\cal N}}\ .
\end{equation}
Clearly, when ${\cal N}$ is large, $\lambda\in(0,1)$.

Also we express the coupling constant $\tilde G$ in terms of
$\alpha$ according to \eqref{eq:alpha_ho}.
Then eq.~\eqref{eq:4.020} becomes
\begin{eqnarray}
\nonumber
&&  \frac{1}{\tilde z}+\frac{1}{\tilde z-1}+\frac{1}{2(\tilde z+1)}+
\frac{1}{2(\tilde z-2)}
  -\log\left|\frac{\tilde z+1}{\tilde z-2}\right|\\
&&  =\frac{1}{\lambda}+\log\frac{1-\lambda}{\lambda}+
  \frac{1-\frac{8}{3}\sqrt{\frac{5}{3}}\alpha}{2\lambda^2}\;,
  \label{eq:4.030}
\end{eqnarray}
the $\tilde z$-dependence appearing only in the lhs, referred to as
$\phi(\tilde z)$, and the $\lambda$ dependence only in the rhs, referred to as
$\xi_\infty(\lambda)$.

\begin{figure}[H]
  \begin{center}
    \epsfig{file=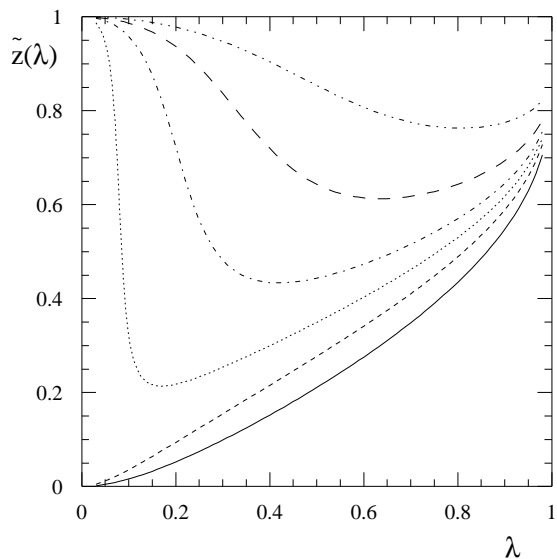,width=10cm,height=10cm}
    \caption{Solution of the eigenvalue equation in the limit 
      ${\cal N}\to\infty$ in the Euler-McLaurin approximation
      for different values of $\alpha$.
      Solid line: $\alpha=0$, dashed line: $\alpha=0.2$, dotted 
      line: $\alpha=0.35$, dash-dotted line: $\alpha=0.5$, long-dashed
      line: $\alpha=0.8$, dot-dot-dashed line: $\alpha=1.5$.
      The exact results are not displayed as they
      almost coincide with the ones in the figure.}
    \label{fig:5}
  \end{center}
\end{figure}

Eq.~\eqref{eq:4.030} is easily solved numerically and the 
results are displayed in Fig.~\ref{fig:5} 
for different values of $\alpha$.

Of significance is the $\alpha\to 0$ case, which still carries the 
fingerprints of the harmonic oscillator and the $\alpha\to\infty$ one, which 
corresponds to the straight line $\tilde z=1$. 
It is of great interest to follow the behaviour with $\alpha$ of
the curves. The sum of the eigenvalues, or, better, of the $\tilde z(\lambda)$,
offers a guidance to globally follow this evolution. Actually this ``sum rule''
for the trapped solutions can be {\it exactly} computed in the two limiting
cases $\alpha=0$ and $\alpha=\infty$. Indeed from the Vi\`ete equations one
has
\begin{equation}
  \label{eq:xxx}
  \Sigma(\alpha=0) 
\equiv \frac{1}{{\cal N}-1} \sum_{k=1}^{{\cal N}-1} \tilde z(k) 
= \frac{1}{4}
\end{equation}
and $\Sigma(\alpha=\infty)=1$, respectively: these values set the limits 
for the area under the curves of Fig.~\ref{fig:5}.

On the other hand the behaviour of the curves themselves
is ruled both by the function  $\xi_\infty(\lambda)$,
which goes to $-\infty$ when $\lambda\to 1$, whereas for $\lambda\to 0$ 
\begin{equation}
  \label{eq:9.303}
  \xi_\infty(\lambda)\underset{\lambda\to0}\longrightarrow
  \begin{cases}
    +\infty&{\rm for ~}\alpha<\alpha_{\rm cr}\\
    -\infty&{\rm for ~}\alpha>\alpha_{\rm cr}
    \end{cases}
\ \ \ \ \ \ \mbox{with}\ \ \ \ 
\alpha_{\rm cr}=\frac{3}{8}\sqrt{\frac{3}{5}}\simeq 0.29\ ,
\end{equation}
and by the monotonic decrease of $\phi(\tilde z)$ 
in the interval $(0,1)$, which varies within the limits
\begin{equation}
  \label{eq:9.305}
  \lim_{\tilde z\to 0}\phi(\tilde z)=
   +\infty\qquad\qquad\lim_{\tilde z\to 1}\phi(\tilde z)=-\infty~.
\end{equation}
Thus, when $\lambda\to1$,
\begin{equation}
\tilde z(\lambda)\to 1+\frac{1}{4\log(1-\lambda)}
\end{equation}
no longer depends upon $\alpha$ and slowly approaches 1. 

For $\lambda\to 0$ two cases occur: if $\alpha<\alpha_{\rm cr}$ 
then $\tilde z(\lambda)\to 0$, if $\alpha>\alpha_{\rm cr}$ instead 
$\tilde z(\lambda)\to 1$. Thus a transition occurs at
$ \alpha_{\rm cr}$: indeed the exact eigenvalue
arises by perturbing the $(k-1)$-th
free one when $\alpha<\alpha_{\rm cr}$ (strong coupling regime), according to
\eqref{eq:4.02}, and by perturbing the $k$-th one when
$\alpha>\alpha_{\rm cr}$ (weak coupling regime) according to
\begin{equation}
  E^{(\nu)}=2e_{\nu}-2 z^{(\nu)}(e_\nu-e_{\nu-1})\ .
\end{equation}

This is strikingly illustrated in Fig.~\ref{fig:5} 
that shows that 
the almost parabolic behaviour of $\tilde z(\lambda)$ for small $\alpha$
is strongly distorted for $\alpha \simeq \alpha_{\rm cr}$ for small $\lambda$.
For larger $\alpha$ a smoother behaviour is recovered. In particular in 
Fig. \ref{fig:5} a marked minimum is seen to develop for $\alpha$
above, but close to, the critical value.
When ${\cal N}$ is finite, for
$\lambda<1/{\cal N}$, there are no eigenvalues, since $\tilde z$ lives on 
a discrete set of points: hence $\alpha_{\rm cr}$ is ill-defined.  

A deeper insight of the above findings is offered by the following comments:
\begin{enumerate}
\item
the pairing interaction is of finite range, therefore a pair trapped by
 highly excited harmonic oscillator states has the
two partners extremely de-localised, hence unaffected by
the interaction. This explains why all 
the curves in Fig.~\ref{fig:5} coalesce to 1 when $\lambda\to 1$, no matter
what the value of $\alpha$ is. On the
other hand all the curves (up to a critical value of $\alpha$) converge to
$\tilde z(0)=0$, reflecting the pressure exercised by the infinite number
of the high-lying, large degeneracy, levels on the low-lying, low-degeneracy, 
ones.
\item
The eigenvalues obey a ``sum rule'', whose value grows with 
$\alpha$ from 1/4 to 1 , hence they must grow as well: since the action 
of the pairing force is gauged by the product $G\Omega_k$, 
at some critical value of $\alpha$ (hence for $G$ sufficiently small)
the system prefers to fulfil the sum rule by 
lifting the lowest eigenvalues (corresponding to the lowest degeneracies)
to the unperturbed values. Indeed (see next section) when the degeneracy of
the harmonic oscillator levels is artificially made to decrease, then no 
transition occurs.
\end{enumerate}

In concluding this section we turn to the problem of finding  
analytical expressions 
for the trapped eigenvalues. In this connection we first notice that, 
remarkably, the large ${\cal N}$ limit is still 
a good approximation to the exact solutions even
for ${\cal N}=5$, as shown in Fig.~\ref{fig:6}.  
While in this limit the complexity of the lhs of eq.~\eqref{eq:4.030}
is still such to render difficult the finding of
simple analytical approximations, yet its structure is dominated by the poles 
in $0$ and $1$ that trap the solution.

It is thus reasonable to 
approximate $\phi(\tilde z)$ with 
\begin{equation}
  \label{eq:4.603}
  \phi^{\rm appr}(\tilde z)=\frac{11}{9}\left(\frac{1}{\tilde z}+
\frac{1}{\tilde z-1}\right)\;,
\end{equation}
which retains the pole structure and reproduces $\phi(\tilde z)$
together with its two first derivatives at $\tilde z=1/2$.
Hence, from the equation \eqref{eq:4.030},
$\phi^{\rm appr}(\tilde z)=\xi_\infty(\lambda)$, it
follows
\begin{equation}
  \label{eq:4.606}
  \tilde z^{\rm appr}(\lambda)=\frac{1}{2}+\frac{11}{9\xi_\infty}
  -\frac{\sqrt{484+81\xi_\infty^2}}{18\xi_\infty}\;,
\end{equation}
displayed as a dotted line in 
Figs.~\ref{fig:6} and \ref{fig:7}.

By expanding in $\lambda$ (say around $1/2$) and in $\alpha$ up to the second
order a parabolic expression for $\tilde z$ as a function of $\lambda$
is obtained, also
shown in Figs.~\ref{fig:6} and \ref{fig:7}.

In these figures the Euler-McLaurin approximation in the case 
${\cal N}\to\infty$  is seen to be remarkably stable and in good accord 
with the exact eigenvalues when $\cal N$ is reduced to finite values 
for $\alpha<\alpha_{cr}$ (but not for $\alpha>\alpha_{cr}$).
Furthermore the predictions of the simple expression
eq.~\eqref{eq:4.606} almost superimpose to the Euler-McLaurin ones.

\noindent

\subsection{Variations on the harmonic oscillator}
\label{sec:4.3}

In this subsection we explore whether the transition previously discussed 
occurs only in the case of the harmonic oscillator or is more general.
Accordingly, we take ${\cal N}$ large and leave
unchanged the regular arrangement of the
harmonic oscillator eigenvalues, but vary, admittedly artificially,
their degeneracy according to  the prescriptions 
\begin{eqnarray}
  \label{eq:4.3.2}
  \Omega_k^{(a)} &=& (k+1)^{\gamma}
 \simeq {\cal N}^{\gamma} \lambda^{\gamma}\\
 \Omega_k^{(b)} &=&
\left({\cal N}-k\right)^\gamma
 \simeq {\cal N}^{\gamma} (1-\lambda)^{\gamma}
\end{eqnarray}
with $\gamma \ge 0$. From the above
\begin{equation}
  \label{eq:4.3.5}
  \alpha^{(a)} = \alpha^{(b)} = \sqrt{\frac{(\gamma+1)^3}{(\gamma+3)}}
\frac{2}{(\gamma+2)\tilde G {\cal N}^{\gamma}}\ .
\end{equation}

Note that for $\gamma$=2 \eqref{eq:4.3.5} yields
$\left.\alpha^{(a)}\right|_{\gamma=2} = \alpha^{(h.o.)}/2$ in the large 
${\cal N}$ limit.

Now from eq.~\eqref{eq:4.07} one gets for
the function $\varphi^{E-McL}(k,\tilde z)$ in the ${\cal N} \to \infty$
limit and in the case (a)  
\begin{equation}
  \label{eq:B1.3}
  \begin{split}
  \varphi^{E-McL}(k,\tilde z)\xrightarrow[{\cal N}\to\infty]{}
  &-\frac{\lambda^\gamma{\cal N}^\gamma}{2}
  \left(\frac{1}{\tilde z+1}+\frac{1}{\tilde z-2}\right)
  -{\cal N}^\gamma\sqrt{\frac{3+\gamma}{(1+\gamma)^3}}(2+\gamma)\alpha^{(a)}\\
 &+{\cal N}^\gamma \left\{\rho(\lambda,\gamma)+
  \lambda^\gamma\log\left|\frac{(1+\tilde z)(1-\lambda)}{(2-\tilde z)
\lambda}\right|  \right\}\ ,
  \end{split}
\end{equation}
(a similar expression holds for case (b)), where 
\begin{equation}
  \label{eq:B1.1}
  \rho(\lambda,\gamma)=
\int\limits_0^1\frac{x^\gamma-\lambda^\gamma}
  {x-\lambda}
\end{equation}
is an {\em analytic} (hence well behaved) function of
$\lambda$ and goes to $1/\gamma$ when $\lambda\to0$.
Accordingly the secular equation can be recast in the form
\begin{eqnarray}
  \label{eq:4.3.7}
  \phi(\tilde z)&=&
   \log \left| \frac{1-\lambda}{\lambda}\right|+
\frac{1}{\lambda^{\gamma}}\left[\rho(\lambda,\gamma)- 
\sqrt{\frac{\gamma+3}{(\gamma+1)^3}}(\gamma+2)\alpha^{(a)}\right]
\end{eqnarray}
in case (a) and
\begin{eqnarray}
  \phi(\tilde z)&=&
 \log  \left|\frac{1-\lambda}{\lambda}\right|-
\frac{1}{(1-\lambda)^{\gamma}}\left[\rho(1-\lambda,\gamma)+ 
\sqrt{\frac{\gamma+3}{(\gamma+1)^3}}(\gamma+2)\alpha^{(b)}\right]\ ,
\nonumber\\
&&\label{eq:4.3.7bis}
\end{eqnarray}
in case (b), the lhs exactly coinciding with the one of eq.~\eqref{eq:4.030}.

In the limits $\lambda \to 0$ (case (a)) and $\lambda \to 1$
(case (b)), the leading terms of \eqref{eq:4.3.7} 
and \eqref{eq:4.3.7bis} are 
\begin{eqnarray} 
&&\frac{1}{\lambda^\gamma}\left[\frac{1}{\gamma}- 
\sqrt{\frac{\gamma+3}{(\gamma+1)^3}}(\gamma+2)\alpha^{(a)} \right]
\label{4.3.71}
\end{eqnarray}
and
\begin{eqnarray}
&&-\frac{1}{(1-\lambda)^\gamma}\left[\frac{1}{\gamma}+
\sqrt{\frac{\gamma+3}{(\gamma+1)^3}}(\gamma+2)\alpha^{(b)}\right]
\label{4.3.71bis}
\end{eqnarray}
respectively.
Remarkably $\tilde z(\lambda)$ in the $\lambda \to 0$
limit goes to 0 or 1, according to the sign of \eqref{4.3.71} (case a), 
which is set by $\alpha$.
The critical value is
\begin{equation}
  \label{eq:4.3.8}
  \alpha_{\rm cr}^{(a)}
  =\sqrt{\frac{(\gamma+1)^3}{\gamma+3}}\frac{1}{\gamma(\gamma+2)} 
\end{equation}
clearly behaving as $1/\gamma$ when 
$\gamma\to 0$. 
On the contrary, in the case (b) in the $\lambda \to 1$
limit $\tilde z(\lambda)$ always tends to 1, 
since \eqref{4.3.71bis} never changes sign. 

Thus a transition occurs only if the degeneracy is growing with $k$ and
the `strong coupling' domain becomes wider ($\alpha_{\rm cr}$
increases) as $\gamma$ approaches zero: here the transition disappears.
If the degeneracy decreases with $k$ (case b), no transition exists.

\begin{figure}[ht]
  \begin{center}
    \epsfig{file=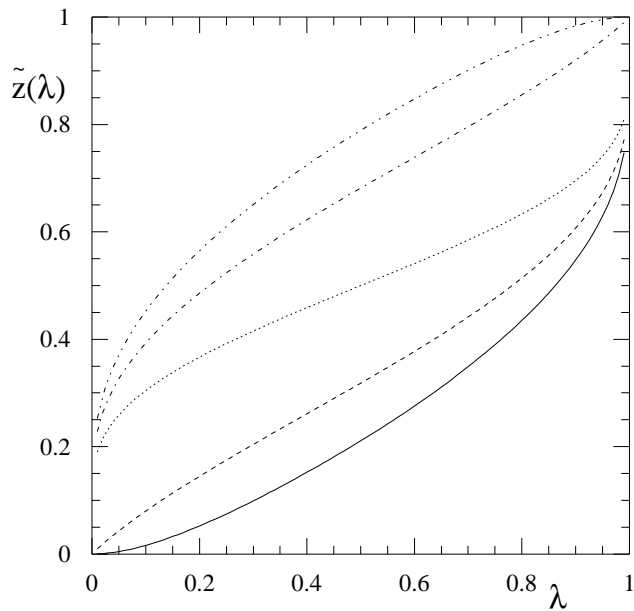,height=10cm,width=10cm}
    \caption{The functions $\tilde z(\lambda)$ for different degeneracies
      and $\alpha=0$. Solid line: case a), $\gamma=2$; dashed line: case a),
      $\gamma=1$; dotted line: case a) and b) $\gamma=0$;
      dash-dotted line: case b), $\gamma=1$, dash-dotted-dotted line:
      case b), $\gamma=2$.}
    \label{fig:8}
  \end{center}
\end{figure}

The eigenvalues corresponding to the degeneracies (a) and (b) for $\alpha=0$ 
are displayed in Fig. \ref{fig:8}.

\section{Conclusions}

In this paper we have once more addressed the pairing problem with the
scopes of finding approximate, but analytic, solutions for the eigenvalues
and eigenvectors  and of disclosing a surprising pattern displayed by the 
former.

To pave the way to the general problem with any number of pairs and levels
we have addressed
the simplified situation with only one pair living in a set of $\cal N$ single
particle levels of a potential well.

For the collective mode we have found, in the strong and weak coupling regimes,
expansions relating its energy not only
to the strength of the pairing interaction, but as well to the parameters
characterising the distribution of the single particle levels.
Thus our analysis shows that different potential wells could lead to the same 
energy for the mode, which is  accurately fixed by only a few parameters of the
levels distribution and not by the precise energy of each of them.
In particular the variance of the levels distribution appears crucial for the
development of a collective mode.

Concerning the energies of the trapped solutions (when these are known, so are
the related eigenfunctions) our search for a simple formula for their
description has in part been prompted by the recent finding \cite{CaRiSa-97} 
concerning the integrability of the pairing Hamiltonian, both at the quantum 
and at the classical level.
Indeed it is reasonable to expect the existence of relatively simple
analytic formulas for the eigenvalues of a Hamiltonian
if the corresponding classical motion is not chaotic.

For this scope the Euler-McLaurin approximation, conveniently
exploited, has been invaluable. 
We started from the extreme case of an infinite number of single particle 
levels of an harmonic oscillator and classified the trapped solutions in terms
of a quantum number $\lambda$, varying between 0 and 1.
For sufficiently large values of the pairing strength $G$ the displacements of
the trapped eigenvalues from the unperturbed solutions always start from 0
and monotonically  grow to 1, reached when $\lambda$=1.
The first of these findings relates to the ``pressure'' exercised 
by the
high-lying eigenvalues (associated with very large degeneracy) on the low-lying
ones, where the degeneracy is low. 
The second one instead relates to the delocalisation of the partners of the 
pair in very high-lying harmonic oscillator levels.
Remember indeed that the pairing interaction is meant to simulate the 
short-range part of the nucleon-nucleon force: hence it is incapable of 
correlating two fermions lying far away from each other, provided no 
quasi-bound is generated by the interaction, as is the case for the 
trapped states.

In this connection it is worth reminding that the classical limit is achieved
by letting the degeneracy of the single particle levels become very large
\cite{RoSiDu-02}. Thus the coincidence of all the eigenvalues, for any $G$, in
$\lambda$=1 also reflects the evolution from quantum to classical
mechanics of our system.

Another aspect of significance of our work concerns the transition of the
behaviour of the eigenvalues from one regime to another. Indeed we 
proved that a critical strength of the pairing force exists such
that for weaker interactions the behaviour of the trapped eigenvalues versus
$\lambda$ ceases to be monotonic. Actually for all $G<G_{crit}$ the 
displacements of the eigenvalues from the unperturbed energies
start from 1 rather than from 0.
This occurrence might be understood on the basis of the sum rule the trapped
solutions should fulfil and of the nature of the pairing force. 
In fact at some point the system prefers to obey the sum rule by lifting
the lower eigenvalues (associated with low degeneracies) to values close to
1 (where the levels are associated with very large degeneracies).

Finally, from our analysis it emerges that the eigenvalues obtained
in the $\cal N\to\infty$ are very robust with
respect to variations of $\cal N$, when $G$ is large: indeed
they keep their validity even for values of $\cal N$ as small as 5.
Moreover, for $G>G_{crit}$, we have 
found that the trapped eigenvalues indeed lend themselves to simple analytical
expressions, even to a parabolic one, as hinted in ref.~\cite{BaFoMoQu-01}.

We are presently investigating the statistical fluctuations of the trapped
eigenvalues, which should reflect the integrability
or, equivalently, the absence of chaotic motion  associated to the pairing
Hamiltonian.

\end{document}